\documentstyle[preprint,aps,epsf,floats,fleqn]{revtex}
\tighten
\begin{document}
\thispagestyle{empty}
\normalsize
\begin{center}
\vspace{3cm}
\large\bf{POLARIZATION OF TAU LEPTONS IN SEMILEPTONIC B DECAYS
\footnote{Work supported in part by KBN grants 2P03B08414 and 
PB659/P03/95/08.} 
} \\
\vspace{1cm}

{\large\bf Marek Je\. zabek$^{a,b}$ \and Piotr Urban$^b$\\
\vspace{0.5cm}
{\normalsize\it   $^a$Institute of Nuclear Physics, Kawiory 26 a,
PL-30055 Cracow, Poland\\
\vspace{0.2cm}
 $^b$Department of Field Theory and Particle Physics, University of 
      Silesia, \\
     Uniwersytecka 4, PL-40007 Katowice, Poland} }
\end{center}
\vspace{1cm}
\begin{abstract}
 Analytic formulae for the $\alpha _s$ order QCD corrections to the 
 differential width of the semileptonic b decay  are given with the 
 $\tau$ polarization taken into account. Thence the polarization of
 $\tau$ is expressed by its energy and the invariant mass of the $\tau +\bar\nu$
 system. The non-perturbative corrections by Falk et al. 
 are incorporated in the calculation.
\end{abstract}
\vspace{3.5cm}
\vspace{1.5cm}
\pagebreak
\setcounter{page}{0}
\footskip 3.0cm
\section{Introduction}
The semileptonic decays of B mesons are now well described with the aid
of the heavy quark effective theory (HQET) \cite{1,2,3,4,5,6} 
and the QCD perturbative corrections.
On the other hand, the nonleptonic processes still suffer from lack
of satisfactory theoretical description \footnote{The accuracy of HQET
for inclusive processes is still under debate, as the assumption of
quark--hadron duality in the final state might introduce $1/m_c$
corrections not seen in the operator product expansion. Phenomenological
analyses suggest that these dangerous terms are absent or small for
two--quark processes like hadronic decays of $\tau$ leptons and
semileptonic decays of heavy quarks, see e.g. \cite{Neubert}
and references therein.}. 
With this in mind, the former 
can be successfully employed in the determination of  the parameters of
the  
Standard Model (SM). Indeed 
they have been used to this end, recently yielding the
Cabibbo-Kobayashi-Maskawa matrix element $|V_{cb}|$ \cite{16}. Moments of lepton
spectra can be used in the determination of $\alpha _s,m_b$ and $m_c$
\cite{13,CJK17,V18}.The particular processes involving the $\tau$ lepton make this field yet
more interesting as they are affected by the mass of the charged lepton,
now comparable with the masses of the involved quarks $m_b$ and $m_c$.
\par 
The first order QCD correction to the differential decay width of the $b$
quark has been found analytically \cite{JL} with the $\tau$ polarization 
summed over.
Now the polarization itself, which is no more suppressed by heavy quark masses,
provides information on the SM parameters in its own right. It does not
depend
on the $|V_{CKM}|$ elements, for instance, but can still yield the quark masses.
This is all the more important due to the fact that the polarization turns out
to be only weakly dependent on the coupling constant $\alpha _s$. It should
be borne in mind that the first order QCD correction to the decay width
itself
is important, amounting to as much as $20\%$ of the Born approximation.
Thus the results of this paper render the $\tau $ polarization especially
applicable for use in evaluation of the quark masses. 
\par
 What we calculate here is the longitudinal polarization of the charged 
lepton. It may be 
added that the longitudinal polarization is transferred to the lepton system
via the intermediating $W^-$ boson whose longitudinal polarization in turn
takes root in the Higgs mechanism, so that the process may shed light on
 physics beyond SM.
\par
We give the formulae for the decay of $b$ quark into the charmed quark, $\tau$
lepton and $\tau$-antineutrino in terms of the charged lepton energy and 
the 
squared four-momentum of the intermediating $W^-$ boson, or, equivalently,
the invariant mass of the $\tau +\bar\nu$ system. These formulae, combined
with the ones for the unpolarized lepton case, readily give the polarization
of the $\tau$ lepton. This expression is then integrated to give $\tau$ energy
distribution. The correction to the polarization is shown together with
the Born-approximated result. 
Then moments of energy distribution are evaluated for the polarized case. 
\par
The sectioning of the paper goes as following. Sec.2 is devoted to the kinematical
variables. In Sec.3 the ideas behind the present calculation of the polarization
are discussed. Then in Sec.4 QCD corrections are briefly described. The final 
analytic result is given in Sec.5 and in the following Sec.6 the moments of
energy distribution are given.

\section{Kinematics}
\subsection{Kinematical variables}
 In this section we define the kinematical variables used throughout the article
as well as the constraints on those in both cases of a 3- and 4-body decay.
 The calculation is performed in the rest frame of the decaying $ b$ quark.
In order to include the first-order QCD corrections to the decay, one must take 
into account both the 3-body final state with a produced quark c, lepton $ \tau$ and
an antineutrino $ {\bar \nu}_{\tau}$ and the 4-body state with an 
additional real gluon.
The four-momenta of the particles are denoted as following: $Q$ for the $b$ quark, 
$q$ for the $c$ quark, $\tau$ for the charged lepton, $\nu$ for the corresponding 
antineutrino and $G$ for the real gluon. All the particles assumed to be on-shell, their
squared four-momenta equal their masses:
\begin{equation}
Q^{2}=m^{2}_{b},\ q^{2}=m^{2}_c,\ \tau^{2}=m^{2}_{\tau},\ \nu^{2}=G^{2}=0.
\end{equation}
The four-vectors P=q+G and W=$\tau+\nu$ characterize the quark-gluon system and the 
virtual intermediating $W$ boson, respectively. The employed variables are scaled in the
units of the decaying quark mass $m_b$:
\begin{equation}\label{variables}
\rho={{m_c^2} \over {m_b^2}},\ \eta={{m_{\tau}^2} \over {m_b^2}},\ 
x={{2E_{\tau}} \over {m_{b}}},\ t={{W^2} \over {m_{b}^{2}}},\ 
z={{P^2} \over {m_{b}^{2}}}.
\end{equation}
Henceforth we scale all quantities so that $m_b^2=Q^2=1$. The charged lepton
 is described by the light-cone variables:
\begin{equation}
\tau_{\pm}={1 \over 2}(x \pm \sqrt{x^{2}-4\eta}).
\end{equation}
 Thus, the system of the $c$ quark and the real gluon is described by the following 
quantities:
\begin{eqnarray}
P_{0}&=&{1 \over 2}(1-t+z)\\
P_{3}&=&\sqrt{P_{0}^{2}-z}={1 \over 2}[1+t^{2}+z^{2}-2(t+z+tz)]^{1 \over 2},\\
P_{\pm}(z)&=&P_{0}(z)\pm P_3(z),\\
{\cal Y}_p&=&{1 \over 2}\ln{{P_+(z)} \over {P_-(z)}}=\ln{{P_+(z)} \over {\sqrt{z}}}
\end{eqnarray}
where $P_0(z)$ and $P_3(z)$ are the energy and the length of the momentum vector
of the system in the $b$ quark rest frame, ${\cal Y}_p(z)$ is the corresponding 
rapidity. Similarly for the virtual boson $W$:
\begin{eqnarray}
W_0(z)={1 \over 2}(1+t-z),\\
W_3(z)=\sqrt{W_0^2-t}={1 \over 2}[1+t^2+z^2-2(t+z+tz)]^{1/2},\\
W_{\pm}(z)=W_0(z)\pm W_3(z),\\
{\cal Y}_w(z)={1 \over 2}\ln{{W_+(z)} \over {W_-(z)}}=\ln{{W_+(z)} \over \sqrt{t}}.
\end{eqnarray}
 Kinematically, the three body decay is a special case of the four body one, 
with the four-momentum of the gluon set to zero, thus resulting in simply
replacing $z=\rho$. The following variables are then useful:
\begin {eqnarray}
p_0=P_0(\rho)={1 \over 2}(1-t+\rho)&,& p_3=P_3(\rho) = 
\sqrt{p_0^2-\rho},\\
p_{\pm}=P_{\pm}(\rho) = p_0\pm p_3&,&  w_{\pm}=W_{\pm}(\rho) = 1-p_{\mp},\\
Y_p={\cal Y}_p(\rho) = {1 \over 2}\ln{{p_+} \over {p_-}}&,&  Y_w={\cal 
Y}_w(\rho) = {1 \over 2}\ln{{w_+} \over {w_-}}. \end{eqnarray}
 We also express the scalar products in terms of the variables used above, so
in the units of the $b$ quark mass one gets:
\begin{eqnarray}
Q\cdot P={1 \over 2}(1+z-t)&,& \tau\cdot\nu ={1 \over 2}(t-\eta),
\nonumber \\
Q\cdot \nu={1 \over 2}(1-z-x+t)&,& \tau \cdot P={1\over 2}(x-t-\eta),
\nonumber \\
Q\cdot \tau={1 \over 2}x&,&\nu\cdot q={1 \over 2}(1-x-z+\eta).
\end{eqnarray}
\subsection{Kinematical boundaries}
 The phase space is divided into two regions. The first of them, 
henceforth called A, is  available for both three- and four-body decay,
while the remaining part B corresponds to pure four-body decay. Region A
is defined as following:
\begin {eqnarray}
2\sqrt{\eta}\le x \le 1+\eta-\rho=x_m,\\
t_1={\tau}_-(1-{\rho \over {1-{\tau}_-}})\le t\le {\tau}_+(1-{\rho \over 
{1-{\tau}_+}})=t_2. \end{eqnarray}
The additional region B of the phase space, where only 4-body decay is allowed,
has the following boundaries:
\begin{equation}
2\sqrt{\eta}\le x\le x_m,\qquad \eta\le t\le t_1.
\end{equation}
Conversely, if x should vary at a fixed value of t, the boundaries read:
\begin{equation}
\eta\le t\le (1-\sqrt{\rho})^2,\qquad w_-+{\eta \over {w_-}}\le x\le 
w_++{\eta \over {w_+}}, \end{equation}
for region A, and
\begin{equation}
\eta\le t\le \sqrt{\eta}(1-{\rho\over{1-\sqrt{\eta}}}),\qquad 
2\sqrt{\eta}\le x\le w_-+{\eta\over{w_-}}. \end{equation}
for region B.
 The upper limit of the mass squared of the $c$-quark--gluon system is in 
both regions given by
\begin{equation}
z_{max}=(1-{\tau}_+)(1-t/{\tau}_+),
\end{equation}
whereas the lower limit depends on the region:
\begin{equation}
z_{min}=\left\{
\begin{array}{ll}
\rho                        & \rm{ in\: Region\: A}\\
(1-{\tau}_-)(1-t/{\tau}_-)  & \rm{ in\: Region\: B}
\end{array}
\right.
\end{equation}

\section{Polarization at tree level}
In order to calculate the longitudinal polarization we must find the differential
decay width for a given polarized final state of the $\tau$ lepton. According
to the definition
\begin{equation}
P={{\Gamma ^+-\Gamma ^-}\over{\Gamma ^++\Gamma^-}}
=1-2{{\Gamma ^-}\over{\Gamma}},
\end{equation}
where $\Gamma=\Gamma ^++\Gamma ^-$. Once we know the width with the polarization
summed over, we only need to find, for instance, the width for the negatively
polarized $\tau$ lepton.
Before going to discuss the QCD corrections to the longitudinal polarization
of the $\tau$ lepton, we stop for a while to look at the tree level situation where 
the calculation is easy to follow. Once we choose the $b$ quark rest frame 
and decide to look for the longitudinal polarization, we can express the lepton 
polarization four-vector $s$ in terms of the four-momenta $Q$ and $\tau$ of the 
$b$ quark and the $\tau$ lepton, respectively:
\begin{equation}\label{stolQ}
s={\cal A}\tau +{\cal B}Q
\end{equation}
This is due to the fact that now only the temporal component of $Q$ does not
vanish, whereas the spatial parts of $s$ and $\tau$ are parallel. The coefficients
${\cal A},{\cal B}$ appearing in the formula above can be evaluated using the 
conditions defining the polarization four-vector $s$:
\begin{eqnarray}
s^2&=&-1\\
s\cdot\tau &=&0.
\end{eqnarray}
Upon this one arrives at the following expressions:
\begin{eqnarray}
{\cal A}^{\pm}&=&\pm {1\over\sqrt{\eta}}{x\over{\tau_+ -\tau_-}},\\
{\cal B}^{\pm}&=&\mp {{2\sqrt{\eta}}\over{\tau_+ -\tau_-}}.
\end{eqnarray}
where the superscripts at ${\cal {A,B}}$ denote the polarization of the lepton.
\par
This observation combines with another one to make the whole calculation simpler.
The total decay width $d\Gamma_0$ at the tree level for the unpolarized case
reads:
\begin{equation}
d\Gamma_0=G_F^2M_b^5|V_{CKM}|^2{\cal M}_{0,3}^{un}d{\cal R}_3(Q;q,\tau,\nu)/\pi^5
\end{equation}
where the matrix element amounts to
\begin{equation}\label{tauliniowy}
{\cal M}_{0,3}^{un}(\tau)= q\cdot\tau Q\cdot\nu. 
\end{equation}
With this kind of linear dependence on the four-momentum $\tau$, it is worth noting
that the matrix element with the $\tau$ polarization taken into account is,
\begin{equation}\label{Kliniowy}
{\cal M}_{0,3}^{pol}={1\over 2}{\cal M}_{0,3}^{un}(K=\tau -ms)={1\over 
2}(q\cdot K) (Q\cdot\nu), \end{equation}
where $m$ stands for the lepton's mass and we have introduced the four-vector $K$
\begin{equation}
K=\tau -ms.
\end{equation}
Applying now the representation (\ref{stolQ}) of the polarization $s$ we readily
obtain the following useful formula for the matrix element with the lepton polarized:
\begin{equation}\label{abexp}
{\cal M}_{0,3}^{\pm}=\mp{{\tau_{\mp}} \over {\tau_+ -\tau_-}}{\cal 
M}_{0,3}^{un}(\tau)\pm{\eta \over{\tau_+ -\tau_-}}{\cal M}_{0,3}^{un}(Q).
\end{equation}
The first term on the right hand side of (\ref{abexp}) can be calculated immediately
once we know the result for the unpolarized case. Thus the problem reduces to
performing this calculation again with the only difference amounting to replacing
the four-momentum $\tau$ of the lepton with that of the decaying quark, $Q$.
The Born-approximated distribution can be written explicitly as
\begin{equation}
{{d\Gamma ^{\pm}}\over{dx}}=12{\Gamma}_0 f_0^{\pm}(x),
\end{equation}
 
 where
\begin{equation}\label{G0}
{\Gamma}_0={{G_F^2m_b^5}\over{192{\pi}^3}}|V_{CKM}|^2.
\end{equation}
The Born level function $f_0(x)$ reads, for the unpolarized case,
\begin{equation}
f_0(x)={1\over 6}\zeta ^2 \tau _3 \left\{ \zeta [x^2-3x(1+\eta)+8\eta]
+(3x-6\eta)(2-x)\right\},
\end{equation}
while the polarized cases are obtained using the function $\Delta f_0$:
\begin{equation}
\Delta f_0(x) = {1\over{12}}\tau _3^2\zeta ^2\{\zeta (3-x-\eta)+3(x-2)\}
\end{equation}
in the following way:
\begin{equation}
f_0^{\pm}(x)={1\over 2}f_0(x)\pm \Delta f_0(x).
\end{equation}
In the formulae above,
\begin{equation}
\tau _3=\sqrt{x^2-4\eta},\qquad \zeta =1-{\rho \over {1-x+\eta}}.
\end{equation}
At the tree level, one can express the polarization integrated over the 
energy of the charged lepton as well:

  \begin{eqnarray}
  P &=& 1 - 1/18\left\{-24\eta^2+(x_m^3-8\eta^{3/2})(3 + \eta - 3\rho) +\right. 
  \nonumber\\&&     12\eta(x_m^2) - 3x_m^4/2+ 3(1 - \eta)^3(\rho-\rho^3/s^4) + 
  \nonumber\\&&      3(-1 + \eta)\rho(1-\rho/s^2)[3(1-\eta)^2 + \rho(3 + 5\eta)]
                     +3T_3S/2+
  \nonumber\\&&      3(x_m-2\sqrt{\eta})(-12\eta - 4\eta^2 + 12\eta\rho - 3\rho^2 + 
                      3\eta\rho^2 + \rho^3) -
  \nonumber\\&&      12\eta\rho^3\ln(s^2/\rho)- 18(2\eta^2 - \rho^2 - \eta^2\rho^2)
                       \ln\left[2\sqrt{\eta}/(x_m+T_3)\right] + 
  \nonumber\\&&     18(1 - \eta^2)\rho^2                    
\left.\ln\left(2s^2\sqrt{\eta}/[(1-\eta)(1-\eta-T_3)- \rho - 
\eta\rho]\right)\right\} 
  \nonumber\\&&/ \left\{T_3S/12+(2\eta^2 - \rho^2 - 
\eta^2\rho^2)\ln[(x_m+T_3 )/(2\sqrt{\eta})] \right.
  \nonumber\\&&   \left. - (1 - 
\eta^2)\rho^2\ln\left\{2\sqrt{\eta}\rho/\left[(1-\eta)(1-\eta+T_3)- 
\rho - \eta\rho\right]\right\} \right\} ,
 \end{eqnarray}
where
\begin{equation}
s=1-\sqrt{\eta},\qquad T_3=\sqrt{x_m^2-4\eta},
\end{equation}
\begin{equation}
S=1-7[(1+\eta)(\eta+\rho ^2)+\rho(1+\eta ^2)]+\eta ^3 +\rho ^3 
+12\eta\rho .
\end{equation}
\par
Anticipating
the subsequent discussion of the QCD corrections, let us already note that the
specific linear dependence as featured in (\ref{tauliniowy},\ref{Kliniowy}) goes
back to the tensorial structure of the matrix element ${\cal M}_{0,3}$:
\begin{equation}\label{MLH}
{\cal M}_{0,3}^{un}(\tau)={\cal L}_{\mu\nu}^{un}{\cal H}^{\mu\nu},
\end{equation}
where ${\cal L}$ and ${\cal H}$ stand for the leptonic and hadronic tensors, respectively.
It is of course the leptonic tensor ${\cal L}$ where the linearity derives 
from: \begin{equation}\label{Lliniowy}
{\cal L}_{\mu\nu}^{\pm}=\mp{{\tau_{\mp}} \over {\tau_+ -\tau_-}}{\cal 
L}_{\mu\nu}^{un}(\tau)\pm{\eta \over{\tau_+ -\tau_-}}{\cal 
L}_{\mu\nu}^{un}(Q). \end{equation}
It is not surprising then that the corrections to the hadronic tensor 
${\cal H}$ will not affect this property.

\section{Calculation of QCD corrections}
The QCD-corrected differential rate for the $b\rightarrow c+{\tau}_-+\bar\nu $ reads:
\begin{equation}
d\Gamma ^{\pm}=d{\Gamma}_0^{\pm}+d{\Gamma}_{1,3}^{\pm}+d{\Gamma}_{1,4}^{\pm},
\end{equation}
where
\begin{equation}
d{\Gamma}_{0}^{\pm}=G_F^2m_b^5|V_{CKM}|^2{\cal M}_{0,3}^{\pm} d{\cal 
R}_3(Q;q,\tau,\nu)/{\pi}^5 \end{equation}
is the Born approximation, while
\begin{equation}
d{\Gamma}_{1,3}^{\pm}={2 \over 3}{\alpha}_sG_F^2m_b^5|V_{CKM}|^2{\cal 
M}_{1,3}^{\pm}d{\cal R}_3(Q;q,\tau,\nu)/{\pi}^6 \end{equation}
comes from the interference between the virtual gluon and Born amplitudes. 
Then, \begin{equation}
d{\Gamma}_{1,4}^{\pm}={2\over 3}{\alpha}_sG_F^2m_b^5|V_{CKM}|^2{\cal 
M}_{1,4}^{\pm}d{\cal R}_4(Q;q,G,\tau,\nu)/{\pi}^7 \end{equation}
is due to the real gluon emission, $G$ denoting the gluon 
four-momentum. $V_{CKM}$ is the Cabibbo-Kobayashi-Maskawa matrix
element corresponding to the $b$ to $c$ or $u$ quark weak transition. The Lorentz 
invariant $n$-body phase space is defined as 
\begin{equation}
d{\cal R}_n(P;p_1,...,p_n)={\delta}^{(4)}(P-\sum{p_i}){\prod}_i{{d^3{\bf p}_i} \over {2E_i}}.
\end{equation}
The superscript $\pm$ refers to the fact that now the polarization of the charged
lepton is taken into account. In order to evaluate the appropriate rates it
is convenient to take advantage of the decomposition (\ref{Lliniowy}) which led to
the formula for the Born-approximated matrix element (\ref{abexp}). As the QCD corrections
influence only the hadronic tensor, the leptonic tensor and thus its linear dependence
on $\tau$ is left intact. This allows us to represent all the involved matrix elements 
in an analogous way and,in fact, the whole decay width is
of the very same form:
\begin{equation}
d\Gamma^{\pm}=\mp{{\tau_+} \over {\tau_+ 
- - - -\tau_-}}d\Gamma^{un}(\tau)\pm{\eta \over{\tau_+ -\tau_-}}d\Gamma^{un}(Q).
\end{equation}
\par
In Born approximation the  contribution to the decay rate into the three-body final state is proportional
to the expression
\begin{eqnarray}
{\cal M}_{0,3}^-&=&{1\over 4}F_0^-(x,t)={1\over 2}q\cdot K 
Q\cdot\nu=\nonumber\\
&=&{{(1-\rho-x+t)}\over {4(\tau _+-\tau _-)}}\left[\tau 
_+(x-t-\eta)-\eta (1+\rho-t)\right] \end{eqnarray}
The three-body phase space is parametrized by Dalitz variables:
\begin{equation}
d{\cal R}_3(Q;q,\tau,\nu)={{{\pi}^2} \over 4}dx dt. 
\end{equation}
The evaluation of the virtual gluon exchange matrix element yields:
\begin{eqnarray}
{\cal M}_{1,3}^{un}(\tau)=-\left[ H_0 q\cdot\tau Q\cdot\nu +H_+\rho Q\cdot\nu Q\cdot\tau +H_- q\cdot\nu q\cdot\tau \right. \nonumber  \\
+{1 \over 2}{\rho}(H_++H_-)\nu\cdot\tau +{1\over 
2}\rho(H_+-H_-+H_L)[\tau\cdot{(Q-q-\nu)}] (Q\cdot\nu)\nonumber\\ 
\left.-{1\over 2}H_L[\tau\cdot{(Q-q-\nu)}] (q\cdot\nu) \right],
\end{eqnarray}
where
\begin{eqnarray}
\nonumber H_0=4(1-Y_pp_0/p_3)\ln{\lambda}_G+(2p_0/p_3)[Li_2(1-{{p_-w_-}\over {p_+w_+}})\\
\nonumber -Li_2(1-{{w_-}\over {w_+}})
- - - -Y_p(Y_p+1)+2(\ln\sqrt{\rho}+Y_p)(Y_w+Y_p)]\\
+[2p_3Y_p+(1-\rho-2t)\ln\sqrt{\rho}]/t+4,\\
H_{\pm}={1\over 2}[1\pm (1-\rho)/t]Y_p/p_3\pm {1\over t}\ln\sqrt{\rho},\\
H_L={1\over t}(1-\ln\sqrt{\rho})+{{1-\rho}\over {t^2}}\ln\sqrt{\rho}+{2\over {t^2}}Y_pp_3+{{\rho} \over t}{{Y_p} \over {p_3}}.
\end{eqnarray}
and then the polarized case requires
\begin{equation}
{\cal M}_{1,3}^-={{\tau _+}\over{\tau _+ -\tau _-}}{\cal M}_{1,3}^{un}(\tau)
- - - -{{\eta}\over{\tau _+ -\tau _-}}{\cal M}_{1,3}^{un}(Q).
\end{equation}
After renormalization, the virtual correction ${\cal M}_{1,3}^{\pm}$ is ultraviolet convergent. However, the infrared 
divergences are left. They are regularized by a small mass of gluon denoted
as ${\lambda}_G$. In accordance with the Kinoshita-Lee-Nauenberg theorem,this
divergence cancels out when the real emission is taken into account.
 The rate from real gluon emission is evaluated by integrating the expression
\begin{equation}
{\cal M}_{1,4}^{un}(\tau)={{{\cal B}_1(\tau)}\over {(Q\cdot G)^2}}-{ {{\cal B}_2(\tau)}\over {Q\cdot G P\cdot G}}+{ {{\cal B}_3(\tau)} \over {(P\cdot G)^2}},
\end{equation}
where
\begin{eqnarray}
{\cal B}_1(\tau)=q\cdot \tau[Q\cdot\nu(Q\cdot G-1)+G\cdot\nu-Q\cdot\nu Q\cdot G],\\
{\cal B}_2(\tau)=q\cdot \tau[G\cdot\nu -q\cdot\nu Q\cdot G+Q\cdot\nu(q\cdot G-Q\cdot G-2q\cdot Q)]\nonumber\\
+Q\cdot\nu(Q\cdot\tau q\cdot G-G\cdot\tau q\cdot Q),\\
{\cal B}_3(\tau)=Q\cdot\nu(G\cdot\tau q\cdot G-\rho\tau\cdot P).
\end{eqnarray}
Taking account of polarization amounts to substituting $K$ for $\tau$ in the 
coefficients ${\cal B}_{1,2,3}$:
\begin{equation}
{\cal M}_{1,4}^-={{\tau _+}\over{\tau _+ -\tau _-}}{\cal 
M}_{1,4}^{un}(\tau)-{{\eta}\over{\tau _+ - \tau _-}}{\cal M}_{1,4}^{un}(Q). 
\end{equation}
The four-body phase space is decomposed as follows:
\begin{equation}
d{\cal R}_4(Q;q,G,\tau,\nu)=dz d{\cal R}_3(Q;P,\tau,\nu)d{\cal R}_2(P;q,G).
\end{equation}
%The four-momentum of the $c$ quark is substituted for by $P-G$ and then the
%integration of ${\cal M}_{1,4}^-$ over $d{\cal R}_2(P;q,G)$ is performed. Lorentz
%invariance allows to reduce all of the appearing integrals to scalar ones:
%\begin{equation}
%I_n=\int{d{\cal R}_2(P;q,g)(Q\cdot G)^n}.
%\end{equation}
%The formulae for $I_n$ have been presented in {CJ}\\
After em\-ploy\-ing the Da\-litz pa\-ra\-metr\-iz\-at\-ion of the three body pha\-se spa\-ce 
${\cal R}_3$
and integration we arrive at an infrared-divergent expression. 
 
%\begin{equation}
%const F_0(x,t)I_{div},
%\end{equation}
%where
%\begin{equation}
%I_{div}=I_{-2}-(1-t+\rho)I_{-1}/(P\cdot G)+\rho I_0/(P\cdot G)^2,
%\end{equation}
The method used in these calculations is the same as the one employed in the previous ones \cite{JK28,CJ29,JL}. The infrared-divergent
part is regularized by a small gluon mass ${\lambda}_G$ which enters into
the expressions as $\ln({\lambda}_G)$. When the three- and four-body contributions
are added the divergent terms cancel out and then the limit ${\lambda}_G \rightarrow 0$
is performed.This procedure yields well-defined double-differential distributions 
of lepton spectra as described below.

\section{Analytical results}
The following formula gives the differential rate of the decay  
$b\rightarrow \tau{\bar 
\nu}X$, $X$ standing for a $c$ quark or a pair of $c$ and a gluon, once the 
lepton is taken to be negatively polarized:  
\begin{equation}\label{formula12}
 {{d{\Gamma}^-}\over{dx\,dt}}=\left\{ 
 \begin{array}{ll}
 12{\Gamma}_0\left[ F^-_0(x,t)-{{2{\alpha}_s}\over{3\pi}}F_{1,A}^-(x,t)\right] & {\rm for\: }(x,t)\:{\rm in\: A}\\
 12{\Gamma}_0{{2{\alpha}_s}\over{3{\pi}}}F_{1,B}^-(x,t) & {\rm for\: }(x,t){\rm \: in\: B}
 \end{array}
 \right.
\end{equation}
$F_{1}^-$ differs according to which region (A or B) it belongs. 
Region A is available for the 3- and 4-body decay, while Region B with a gluon only.
The following formulae are given for the negative polarization of the lepton,
that is, we take 
\begin{eqnarray}
{\cal A}^{-} &=&- {1\over\sqrt{\eta}}{x\over{\tau_+ -\tau_-}},\\
{\cal B}^{-}&=& {{2\sqrt{\eta}}\over{\tau_+ -\tau_-}}.
\end{eqnarray}
The factor $\Gamma _0$ is defined in Eq.(\ref{G0}), while

\begin{equation}
F_0^-(x,t)=
{{(1-\rho-x+t)}\over {\tau _+-\tau _-}}\left[\tau _+(x-t-\eta)-\eta 
(1+\rho-t)\right] 
\end{equation}
and
\begin{eqnarray}\label{F1main}
 F_{1,A}^-(x,t)= F_0^-{\Phi}_0+{\sum_{n=1}^5 D_n^A{\Phi}_n+D_6^A},\\
 F_{1,B}^-(x,t)= F_0^-{\Psi}_0+{\sum_{n=1}^5 D_n^B{\Psi}_n+D_6^B}.
\end{eqnarray}
The factor 12 in the formula (\ref{formula12}) is introduced to meet the widely used 
\cite{CJK17,FLNN22,V18} convention for
$F_0(x)$ and ${\Gamma}_0$. The symbols present in (\ref{F1main}) are defined as follows:
\begin{eqnarray}\label{Phis}
{\Phi}_0={{2p_0}\over{p_3}}[Li_2(1-{{1-{\tau}^+}\over{p_+}})
+Li_2(1-{{1-t/{\tau}^+}\over{p_+}})\nonumber\\
- - - -Li_2(1-{{1-{\tau}^+}\over{p_-}})-Li_2(1-{{1-t/{\tau}^+}\over{p_-}})\nonumber\\
+Li_2(w_-)-Li_2(w_+)+4Y_p\ln\sqrt{\rho}]\nonumber\\
+4(1-{{p_0}\over{p_3}}Y_p)\ln(z_{max}-\rho)-4\ln z_{max},\\
{\Phi}_1=Li_2(w_-)+Li_2(w_+)-Li_2({\tau}_+)-Li_2(t/{\tau}_+)\\
{\Phi}_2={{Y_p}\over{p_3}},\\
{\Phi}_3={1\over 2}\ln\sqrt{\rho},\\
{\Phi}_4={1\over 2}\ln(1-{\tau}_+),\\
{\Phi}_5={1\over 2}\ln(1-t/{\tau}_+),\\
{\Psi}_0=4({{p_0}\over{p_3}}Y_p-1)\ln({{z_{max}-\rho}\over{z_{min}-\rho}})+4\ln({{z_{max}}\over{z_{min}}})\nonumber\\
+{{2p_0}\over{p_3}}[Li_2(1-{{1-{\tau}_+}\over{p_-}})+Li_2(1-{{1-t/{\tau}_+}\over{p_-}})\nonumber\\
- - - -Li_2(1-{{1-{\tau}_+}\over{p_+}})-Li_2(1-{{1-t/{\tau}_+}\over{p_+}})\nonumber\\
+Li_2(1-{{1-{\tau}_-}\over{p_+}})+Li_2(1-{{1-t/{\tau}_-}\over{p_+}})\nonumber\\
- - - -Li_2(1-{{1-{\tau}_-}\over{p_-}})-Li_2(1-{{1-t/{\tau}_-}\over{p_-}})],\\
{\Psi}_1=Li_2({\tau}_+)+Li_2(t/{\tau}_+)-Li_2({\tau}_-)-Li_2(t/{\tau}_-),\\
{\Psi}_2={1\over 2}\ln(1-{\tau}_-),\\
{\Psi}_3={1\over 2}\ln(1-t/{\tau}_-),\\
{\Psi}_4={1\over 2}\ln(1-{\tau}_+),\\
{\Psi}_5={1\over 2}\ln(1-t/{\tau}_+).
\end{eqnarray}
We introduce $C_1...C_5$ to simplify the formulae for $D_n^A$ and $D_n^B$:
\begin{eqnarray}
C_1 &=& {1\over\sqrt{x^2-4\eta}} \left[\rho (-\eta x+4 \eta t-\eta {\tau}_+-4 \eta+x {\tau}_+-t {\tau}_+)-2 \rho^2 \eta\right.
     +\eta (x t+x   \nonumber\\
&&     +4 t-2 t^2+6+2 \eta)
     \left.    +{\tau}_+ (-\eta x+\eta t+5 \eta-2 x t+x+t+t^2)\right] ;
\end{eqnarray}
\begin{eqnarray}
C_2 &=& \rho (\eta-t-2 {\tau}_+)+ (x t^2-2 t^2 {\tau}_+)/(2\eta)+ \eta (-x-2 t+2 {\tau}_++30)/2  \nonumber\\
&&+t (1+t-x)+{\tau}_+ (-2 x+2 t+6);\\
C_3 &=& {1\over\sqrt{x^2-4\eta}} \left[\rho (-10 \eta x+18 \eta t-7 \eta {\tau}_++x t+7 x {\tau}_+-11 t {\tau}_++4 {\tau}_+)\right.\nonumber\\
&&+\rho^2 (-9 \eta+{\tau}_+)
       +(2 x t^2 {\tau}_+-x^2 t^2/2-t^2 {\tau}_+^2)/\eta+\eta^2 (-{\tau}_++10)
\nonumber\\
&&       +x t (x-t-1)
                \eta (4 x t-x {\tau}_+-2 x-x^2/2+t {\tau}_++4 t\nonumber\\
&&                -4 t^2+11 {\tau}_+-{\tau}_+^2-12)+
      \left.    +{\tau}_+ (-6 x t-x+2 x^2-3 t+5 t^2-5)\right];\\
C_4 &=& {1\over\sqrt{x^2-4\eta}} \left[\rho (2 \eta x 
{\tau}_+/t-8 \eta x+8 \eta {\tau}_+/t+18 \eta t-11 \eta {\tau}_+
- - - -4 \eta^2 {\tau}_+/t^2\right.\nonumber\\
&&     -x t+5 {\tau}_+ x-7 t {\tau}_+)+\rho^2 (2 \eta {\tau}_+/t-9 \eta-\eta^2 {\tau}_+/t^2)+x t (1+t-x)\nonumber\\
&&                +(-2 x t^2 {\tau}_++x^2 t^2/2+t^2 {\tau}_+^2)/\eta+\eta^2 (5 {\tau}_+/t^2-6 {\tau}_+/t-16/t+6)\nonumber\\
&&                +\eta (2 x {\tau}_+/t+2 x t-3 x {\tau}_+-4 x+x^2/2-10 {\tau}_+/t+3 t {\tau}_++12 t\nonumber\\
&&  \left.               -4 t^2+3 {\tau}_++{\tau}_+^2+4)+{\tau}_+ (-4 x t-3 x+2 x^2+11 t+2 t^2) \right];\\
C_5&=& [\rho /(x-t-\eta/t)] (-\rho \eta^2 {\tau}_+/t^2+\eta {\tau}_+/t-\eta-\eta^2 {\tau}_+/t^2+\eta^2/t+\rho \eta {\tau}_+/t\nonumber\\
&&-\rho \eta
   +\rho \eta^2/t)/2+ [\rho /(1-x+\eta)] (-\eta t-\eta {\tau}_++\eta^2+t {\tau}_+)/2\nonumber\\
&&      +\rho (3 \eta {\tau}_+/t+9 \eta-9 t-3 {\tau}_+)/2+(x t^2/4-t^2 {\tau}_+)/\eta+\eta (3 x-10 {\tau}_+/t\nonumber\\
&&-12 t-2 {\tau}_+-24+2 \eta)/4+(3 t+5) {\tau}_+/2-x t+6 t+5 t^2/2;     \\
D^A_1&=&C_1;\\
D^A_2&=& {1\over{4 \sqrt{x^2-4\eta}}}\left\{\rho \eta (34-6 x {\tau}_+/t^2-4 x/t^2+3 x/t+5 x t-2 x {\tau}_++2 x
 \right.\nonumber\\
&& +3 x^2/t^2
 +x^2+8 {\tau}_+/t^2+18 {\tau}_+/t-2/t+2 t {\tau}_++38 t-14 t^2+20 {\tau}_+)\nonumber\\
&& +\rho \eta^2 (16-2 x/t^2-5 {\tau}_+/t^2-2 {\tau}_+/t+16/t-{\tau}_+)+\rho {\tau}_+ (-4 x t-4 x+6 t\nonumber\\
&&+7 t^2+11)
   +\rho^2 \left[\eta (-10+6 x {\tau}_+/t^2+6 x/t^2+4 x {\tau}_+/t-3 x-3 x^2/t^2\right.\nonumber\\
&&   -2 x^2/t-12 {\tau}_+/t^2
  -18 {\tau}_+/t+6/t+18 t-6 {\tau}_+)+\eta^2 (x/t^2+{\tau}_+/t^2\nonumber\\
&&-{\tau}_+/t-2/t)
  +{\tau}_+ (2 x-5 t-7)+
          +\rho \eta (-10-2 x {\tau}_+/t^2-4 x/t^2-x/t\nonumber\\
&&+x^2/t^2+8 {\tau}_+/t^2
          +6 {\tau}_+/t-6/t)+\rho \eta^2 {\tau}_+/t^2 
 +\rho {\tau}_++\rho^2 \eta (x/t^2-2 {\tau}_+/t^2\nonumber\\
&&\left.+2/t)\right]+{\tau}_+ (-4 x t+2 x t^2+2 x+7 t+t^2-3 t^3-5)
   +\eta (-14+2 x {\tau}_+/t^2\nonumber\\
&& +x/t^2-4 x {\tau}_+/t-2 x/t+4 x t-2 x t^2+2 x {\tau}_+-x-x^2/t^2
        +2 x^2/t-x^2\nonumber\\
&&-2 {\tau}_+/t^2
- - - -6 {\tau}_+/t-10 t {\tau}_++32 t-22 t^2+4 t^3+18 {\tau}_+)
    +\eta^2 (28+x/t^2\nonumber\\
&&-2 x/t
 \left.+x+3 {\tau}_+/t^2-5 {\tau}_+/t-14/t+t {\tau}_+-14 t+{\tau}_+)\right\};\\
D^A_3 &=&{1\over\sqrt{x^2-4\eta}}\left\{\rho \left[\eta (-6-4 x {\tau}_+/t^2-3 x/t^2-2 x {\tau}_+/t+x/t-13 x+2 x^2/t^2\right.\right.\nonumber\\
&&  +x^2/t+6 {\tau}_+/t^2+12 {\tau}_+/t-2/t+26 t-10 {\tau}_+)+\eta^2 
(-x/t^2-2 {\tau}_+/t^2\nonumber
\\&&+2/t)  \left.+{\tau}_+ (6 x+6-12 t)\right]+\rho^3 \eta (-x/t^2+2 {\tau}_+/t^2-2/t)\nonumber\\
&&  +\rho
^2 \left[\eta (-10+2 x {\tau}_+/t^2+3 x/t^2-x^2/t^2-6 
{\tau}_+/t^2-4 {\tau}_+/t+4/t)\right.\nonumber\\
&&  \left.-\eta^2 {\tau}_+/t^2-{\tau}_+\right]+\eta^2 (14+x/t^2-x/t+3 {\tau}_+/t^2-2 {\tau}_+/t-14/t-{\tau}_+)\nonumber\\
&&  +\eta (-14+2 x {\tau}_+/t^2+x/t^2-2 x {\tau}_+/t-x/t+2 x t-2 x-x^2/t^2+x^2/t\nonumber\\
&&  \left.-2 {\tau}_+/t^2-8 {\tau}_+/t+18 t-4 t^2+10 {\tau}_+)+{\tau}_+ (-2 x t+2 x+2 t+3 t^2-5)\right\};  \nonumber\\
\\
D^A_4&=&-C_3-C_2;\\
D^A_5&=&-C_4+C_2;\\
D^A_6&=&-{1\over 2}C_5+{1\over{4 \sqrt{x^2-4\eta}}} \left\{[\rho /(1-x+\eta)]  \left[\eta (t {\tau}_++t+{\tau}_+)-\eta^2 (1+t+{\tau}_+)\right.\right.\nonumber\\
&&\left.+\eta^3-t {\tau}_+)\right]
 +[\rho /(x-t-\eta/t)] (1+\rho) \left[\eta (t-{\tau}_+)+\eta^2 (-1+{\tau}_+/t^2+\right.\nonumber\\
&&  {\tau}_+/t -1/t)
 \left.+\eta^3 (1/t^2-{\tau}_+/t^3)\right]+\rho \eta (43-x {\tau}_+/t-4 x/t-5 x+2 x^2/t\nonumber\\
&& +9 {\tau}_+/t  +13 t+5 {\tau}_+)
 +\rho \eta^2 (-5-{\tau}_+/t^2+2 {\tau}_+/t+1/t)+{\tau}_+ (9 x t+5 x \nonumber\\
&& +4 t-6 t^2) +12 x t+5 x t^2-2 x^2 t
 +\rho (-9 x t-3 x {\tau}_++5 t {\tau}_+) +\nonumber\\
&&  \rho^2 \eta (-10+2 x/t   -3 {\tau}_+/t)
 +\rho^2 \eta^2 (-{\tau}_+/t^2+1/t)+ (-6 x t^2 {\tau}_++x^2 t^2 \nonumber\\
&& +2 t^2 {\tau}_+^2)/(2\eta)
+\eta (-35-x {\tau}_+/t+2 x/t-2 x t+4 x {\tau}_++26 x \nonumber\\
&&   -2 x^2/t-3 x^2/2-4 {\tau}_+/t
- - - -2 t {\tau}_+-34 t-t^2-22 {\tau}_+-3 {\tau}_+^2)  \nonumber\\
&& \left. +\eta^2 (-2+2 x/t-x+6 {\tau}_+/t+2 t)\right\}; 
\end{eqnarray}

\begin{eqnarray}
 D_1^B &=&  C_1\\
 D_2^B &=&  C_2 -  C_3\\
 D_3^B &=& -C_2 -  C_4\\
 D_4^B &=&  C_2 +  C_3\\
 D_5^B &=& -C_2 +  C_4\\ 
 D_6^B &=&  C_5\\
\end{eqnarray}
 One can perform the limit $\rho\rightarrow 0$, which corresponds to the decay
 of the bottom quark to an up quark and leptons. The formulae, which are much 
 simpler in this case, are presented in the same manner as the full results:
\begin{equation}
{{ d\widetilde{\Gamma}^-}\over{dx\,dt}}=\left\{
 \begin{array}{ll}
 12{\Gamma}_0\left[ 
{\widetilde{F}}_0^-(x,t)-{{2{\alpha}_s}\over{3\pi}}{\widetilde{F}}_{1,A}^-(x,t)\right] & \rm for (x,t) in A\\
 12{\Gamma}_0{{2{\alpha}_s}\over{3{\pi}}}{\widetilde{F}}_{1,B}^-(x,t) & \rm for (x,t) in B
 \end{array}
 \right.
\end{equation}
with 
\begin{equation}
\widetilde{F}_0^-(x,t)=
{{(1-x-t)}\over {\tau _+-\tau _-}}\left[\tau _+(x-t-\eta)-\eta (1-t)\right]
 \end{equation}
and
\begin{eqnarray}
\widetilde{F}_{1,A}^-(x,t)=\widetilde{F}_0{\widetilde{\Phi}}_0+{\sum_{n=1}^5 {\cal D}_n^A{\widetilde{\Phi}}_n+{\cal D}_6^A},\\
\widetilde{F}_{1,B}^-(x,t)=\widetilde{F}_0{\widetilde{\Psi}}_0+{\sum_{n=1}^5 {\cal D}_n^B{\widetilde{\Psi}}_n+{\cal D}_6^B}.
\end{eqnarray}
where
\begin{eqnarray}
{\widetilde{\Phi}}_0=2\left[Li_2({{{\tau}_+-t}\over{1-t}})+Li_2({{1/{{\tau}_+}-1}\over{1/t-1}})+Li_2(t)\right]+{1\over 2}{\pi}^2\nonumber\\
\nonumber +\ln ^2(1-{\tau}_+)_2\ln ^2(1-t)+\ln ^2(1-t/{{\tau}_+})\\
 -2\ln (1-t)\ln z_{max},\\ \nonumber
{\widetilde {\Phi}}_1={{\pi ^2}\over 12}+Li_2(t)-Li_2(\tau _+)-Li_2(t/{\tau _+}),\\
{\widetilde {\Phi}}_{2\Diamond 3}={{2\ln (1-t)}\over {1-t}},\\ 
{\widetilde {\Phi}}_4={\Phi}_4,\\ 
{\widetilde {\Phi}}_5={\Phi}_5,
\end{eqnarray}
and
\begin{eqnarray}
{\widetilde{\Psi}}_0=2\left[ Li_2({ {{\tau_+}-t} \over{1-t} })+Li_2({ {{1/{\tau_-}}-1} \over{1/t-1}})-Li_2({ {{\tau_+}-t}\over {1-t} })\right. \nonumber\\
\left.-Li_2({{1/{{\tau}_+}-1}\over {1/t-1}})\right]+\ln (1-t)\ln ({{z_{max}}\over{z_{min}}})\nonumber\\ \nonumber
- - - -\ln ({{1-{{\tau}_+}}\over{1-{{\tau}_-}}})\ln \left[ (1-{{\tau}_+})(1-{{\tau}_-})\right] \\ 
- - - -\ln ({{1-t/{{\tau}_+}}\over{1-t/{{\tau}_-}}})\ln \left[ (1-t/{{\tau}_+})(1-t/{{\tau}_-})\right],\\ 
{\widetilde{\Psi}}_n={\Psi}_n, \qquad  (n=1...5).
\end{eqnarray}
\begin{eqnarray}
{\cal C}^A_1&=&{1\over\sqrt{x^2-4\eta}} \left[\eta (6+x t-x {\tau}_++x+t {\tau}_++4 t-2 t^2+5 {\tau}_+)+2 \eta^2+\right.
\nonumber\\&&              \left.    {\tau}_+(x-2 x t +t +t^2 )\right];
\end{eqnarray}
\begin{eqnarray}
{\cal C}^A_2&=& (x t^2-2 t^2 {\tau}_+)/(2\eta)+ \eta (30-x-2 t+2 {\tau}_+)/2-x t-2 x {\tau}_++2 t {\tau}_+
\nonumber\\&&   +t+t^2+6 {\tau}_+;   \\
{\cal C}^A_3&=&{1\over\sqrt{x^2-4\eta}} \left[(4 x t^2 {\tau}_+-x^2 t^2-2 t^2 {\tau}_+^2)/(2\eta)+
+\eta^2 (10-{\tau}_+)    \right. \nonumber   \\
&&              \eta (-24+8 x t-2 x {\tau}_+-4 x-x^2+2 t {\tau}_++8 t-8 t^2+22 {\tau}_+-2 {\tau}_+^2)/2 \nonumber    \\
&&     \left.   -6 x t {\tau}_+-x t-x t^2-x {\tau}_++x^2 t+2 x^2 {\tau}_+
- - - -3 t {\tau}_++5 t^2 {\tau}_+-5 {\tau}_+\right];     \\
{\cal C}^A_4&=&{1\over\sqrt{x^2-4\eta}} \left[  (-4 x t^2 {\tau}_+
+x^2 t^2+2 t^2 {\tau}_+^2)/(2\eta)+\eta (4+2 x {\tau}_+/t   +2 x t       \right.  \nonumber  \\
&&-3 x {\tau}_+-4 x+x^2/2    -10 {\tau}_+/t+3 t {\tau}_++12 t-4 t^2+3 {\tau}_++{\tau}_+^2) \nonumber \\
&&  +\eta^2 (6+5 {\tau}_+/t^2-6 {\tau}_+/t
- - - -16/t)  -4 x t {\tau}_++x t+x t^2-3 x {\tau}_+  \nonumber \\
&& \left.  -x^2 t+2 x^2 {\tau}_++11 t {\tau}_++2 t^2 {\tau}_+\right];    \\
{\cal C}^A_5&=& (x t^2-4 t^2 {\tau}_+)/(4\eta)+ \eta (-24+3 x-10 {\tau}_+/t-12 t-2 {\tau}_+)/4 \nonumber
\\&&+\eta^2/2-x t+3 t {\tau}_+/2+6 t+5 t^2/2+5 {\tau}_+/2;    \\
{\cal D}^A_1&=&{\cal C}_1    \\
{\cal D}^A_{2\Diamond 3}&=&{1\over{4\sqrt{x^2-4\eta}}} 
\left[ \eta (-14+2 x {\tau}_+/t^2+x/t^2-4 x {\tau}_+/t-2 x/t+4 x t-2 x t^2\right.  \nonumber
\\&&    +2 x {\tau}_+-x-x^2/t^2+2 x^2/t-x^2-2 {\tau}_+/t^2-6 {\tau}_+/t-10 t {\tau}_++32 t  \nonumber
\\&&-22 t^2+4 t^3
      +18 {\tau}_+)+ \eta^2 (28+x/t^2-2 x/t+x+3 {\tau}_+/t^2-5 {\tau}_+/t  \nonumber
\\&&-14/t
      +t {\tau}_+
   \left.-14 t+{\tau}_+)+ (-4 x t {\tau}_++2 x t^2 {\tau}_++2 x 
{\tau}_++7 t {\tau}_++t^2 {\tau}_+\right.  \nonumber
\\&&-3 t^3 {\tau}_+
 \left.   -5 {\tau}_+)\right];    \\
{\cal D}^A_4&=&-{\cal C}_2-{\cal C}_3,   \\
{\cal D}^A_5&=&{\cal C}_2-{\cal C}_4,    \\
{\cal D}^A_6&=&-{1\over 2}{\cal C}_5+{1\over{4\sqrt{x^2-4\eta}}} 
\left[ (-6 x t^2 {\tau}_++x^2 t^2+2 t^2 {\tau}_+^2)/(2\eta)\right.  \nonumber\\
&&     + \eta (-35-x {\tau}_+/t+2 x/t-2 x t+4 x {\tau}_++26 x-2 x^2/t-3 x^2/2-4 {\tau}_+/t  \nonumber\\
&&     -2 t {\tau}_+-34 t-t^2-22 {\tau}_+-3 {\tau}_+^2)  
      + \eta^2 (-2+2 x/t-x+6 {\tau}_+/t+2 t) \nonumber\\
&&     \left. + (9 x t {\tau}_++12 x t+5 x t^2+5 x {\tau}_+-2 x^2 t+4 t {\tau}_+-6 t^2 {\tau}_+)\right];
\end{eqnarray}
\begin{figure}[htb]
\begin{center}
\leavevmode
\epsfxsize = 270pt
\epsfysize = 270pt
\epsfbox[1 140 578 702]{fig1.ps}
\end{center}
\caption{Polarization of $\tau$ lepton in the Born approximation (dashed line)
 and including the first order QCD correction (solid line) 
 as functions of the scaled $\tau$ energy x. The mass of the $b$ quark
 taken at $4.75$ GeV, $c$ quark $1.35$ GeV and the coupling constant $\alpha _s=0.2$}
\label{FIG1}
\end{figure}
 On integration over $t$, one obtains tau energy distributions according 
to the formula
\begin{equation}\label{DEFF1}
{1 \over {12{\Gamma}_0}}{{d\Gamma ^-}\over{dx}}=f_0^-(x)-{{2{\alpha}_s} \over {3\pi}}f_1^-(x).
\end{equation}
 where if we drop the superscript '$-$' we obtain the corresponding formula
for the unpolarized case.
\par
 The results are presented in Fig. \ref{FIG1}, where the polarization
is plotted versus the $\tau$ lepton energy. Both the Born and first order 
approximation are showed. 
In order to make the correction more explicit, we
also present another diagram, where the function $R(x)$ is drawn, defined
as following:
\begin{equation}\label{PviaR}
1-P(x)=[1-P_0(x)][1+{{2\alpha _s}\over {3\pi}}R(x)]
\end{equation}
which gives the expression for $R(x)$:
\begin{equation}
R(x)={{f_1(x)}\over{f_0(x)}}-{{f_1^-(x)}\over{f_0^-(x)}}
\end{equation}
thus its meaning is how the radiative corrections differ with respect to
the state of polarization they act on. In (\ref{PviaR}) $P_0$ denotes the
zeroth order approximation to the polarization. The function $f_1^-(x)/f_0^-(x)$ has 
been presented in Fig. \ref{FIG2}, while Fig. \ref{FIG3} shows the
function $R(x)$, so that
one can immediately see how small the correction is in comparison to the
correction to the energy distribution.
\begin{figure}[htb]
\begin{center}
\leavevmode
\epsfxsize = 250pt
\epsfysize = 250pt
\epsfbox[1 140 578 702]{fig2.ps}
\end{center}
\caption{ The ratio $f_1^-(x)/f_0^-(x)$ representing the radiative
correction for the negatively polarized state as dependent on the scaled
$\tau$ lepton energy x.}
\label{FIG2}
\end{figure}
%%%
\begin{figure}[htb]
\begin{center}
\leavevmode
\epsfxsize = 250pt
\epsfysize = 250pt
\epsfbox[1 140 578 702]{fig3.ps} \vspace{1em} 
\end{center}
\caption{ The QCD-correction 
function R(x)  for the pole mass values of 
the $b$ quark taken to be $4.4$ GeV (dashed), $4.75$ GeV (solid) and 
$5.2$ GeV (dash-dotted) as dependent on the scaled $\tau$ lepton energy x.}
\label{FIG3}
\end{figure}

\par
 Integrating over the charged lepton energy one can obtain its total 
polarization as well as the corrections to which it is subject. If we 
take $m_b=4.75$ GeV as the central value for the decaying quark mass and 
$m_b=4.4$ and $m_b=5.2$ for the limits, we arrive at the following (the 
mass difference $m_b - m_c=3.4$ GeV everywhere): 
\begin{equation}
1-P=(1-P_0)\left( 1+{{2\alpha _s}\over{3\pi}}R_s +{{\lambda _1^2}\over 
{m_b^2}}R_{np}^1 + {{\lambda _2^2}\over{m_b^2}}R_{np}^2\right)
\end{equation}
with
\begin{eqnarray*}
P_0&=&-0.7388^{-0.0105}_{+0.0109}\\
R_s&=&-0.016^{-0.023}_{+0.017}\\
R_{np}^1&=&\ 0.421^{+0.027}_{-0.025}\\
R_{np}^2&=&-2.28^{-0.22}_{+0.16}\\
\end{eqnarray*}

\section{Moments of $\tau$ energy distribution}
 The moments of $\tau$ energy distribution, which are useful sources of information
on the physical parameters regarding the discussed decay  can be evaluated according
to the formula:
\begin{eqnarray}
M_n^{\pm}=\int_{E_{min}}^{E_{max}}E^n_{\tau}{{d\Gamma 
^{\pm}}\over{d{E_{\tau}}}}dE_{\tau},\\
r_n^{\pm}={{M_n^{\pm}}\over{M_0^{\pm}}},
\end{eqnarray}
where $E_{min}$ and $E_{max}$ are the lower and upper limits for $\tau$ energy and
$M_n$ include both perturbative and nonperturbative QCD corrections to $\tau$ energy
spectrum. The superscripts denote the polarization states. Since one obviously 
has 
\begin{equation}
M_n=M_n^++M_n^-
\end{equation}
where $M_n$ stands for the unpolarized momenta, we only give the values of
the momenta for the negative polarization case. The unpolarized distributions
were given in \cite{JL}. The nonperturbative corrections to the charged lepton spectrum from
semileptonic $B$ decay have been derived in the HQET framework  
up to order of $1/m_b^2$ \cite{FLNN22,K23,BKPS24}. The corrected heavy lepton
energy spectrum can be written in the following way:
\begin{equation}\label{HQET1}
{1 \over {12{\Gamma}_0}}{{d\Gamma}\over{dx}}=f_0(x)-{{2{\alpha}_s} \over {3\pi}}f_1(x)+{{{\lambda}_1}\over{m_b^2}}f_{np}^{(1)}(x)+{{{\lambda}_2}\over{m_b^2}}f^{(2)}_{np}(x),
\end{equation}
where ${\lambda}_1$ and ${\lambda}_2$ are the HQET parameters corresponding to the $b$
quark kinetic energy and the energy of interaction of the $b$ quark magnetic moment
with the chromomagnetic field produced by the light quark in the meson $B$. The functions
$f_{np}^{(1,2)}$ can be easily extracted from the formula (2.11) in \cite{FLNN22}.The 
formula (\ref{HQET1}) looks identically if one considers definite polarization
state of the final $\tau$ lepton. The appropriate calculation within the
HQET scheme has also been performed \cite{FLNN22}, see formula (2.12) therein.
\par
  Following ref.\cite{V18} we expand the ratios $r_n$:
\begin{equation}
r_n^-=r_n^{(0)-}\left(1-{{2{\alpha}_s}\over {3\pi}}{\delta}_n^{(p)-}+{{\lambda _1}\over{m_b^2}}{\delta}_n^{(1)-}+{{\lambda _2}\over {m_b^2}}{\delta}_n^{(2)-}\right),
\end{equation}
where $r_n^{(0)}$ is the lowest approximation of $r_n$,
\begin{equation}
r_n^{(0)-}=({{m_b}\over 2})^n{{\int_{2\sqrt{\eta}}^{1+\eta 
- - - -\rho}f_0^-(x)x^ndx}\over{\int_{2\sqrt{\eta}}^{1+\eta -\rho}f_0^-(x)dx}}.
\end{equation}
Each of the ${\delta}_n^{(i)-}$ is expressed by integrals of the corresponding correction
function $f^{(i)}(x)$ and the tree level term $f_0(x)$
\begin{equation}
{\delta}_n^{(i)-}={{\int_{2\sqrt{\eta}}^{1+\eta 
- - - -\rho}f^{(i)-}(x)x^ndx}\over{\int_{2\sqrt{\eta}}^{1+\eta 
- - - -\rho}f_0^-(x)x^ndx}}-{{\int_{2\sqrt{\eta}}^{1+\eta 
- - - -\rho}f^{(i)-}(x)dx}\over{\int_{2\sqrt{\eta}}^{1+\eta -\rho}f_0^-(x)dx}},
\end{equation}
where the index $i$ denotes any of the three kinds of corrections discussed
above. The coefficients $\delta _n^{(i)}$~ depend only on the two ratios
of the charged lepton and the $c$~ quark to the mass $m_b$. Following 
ref.\cite{JL} we employ the functional dependence of the form
\begin{equation}
\delta^{(i)-}_n(m_b,m_c,m_{\tau})=\delta^{(i)-}_n\left({{m_b}\over{m_{\tau}}},
{{m_c}\over{m_b}}\right).
\end{equation}
The quark masses are not known precisely so we have calculated the coefficients
in a reasonable range of the parameters, that is, $4.4$~GeV$\leq m_b \leq 5.2$~
GeV and $0.25 \leq m_c/m_b \leq 0.35$ and then fitted to them functions of
the following form:
\begin{eqnarray}\label{deltapq}
\delta (p,q)&=&a+b(p-p_0)+c(q-q_0)+d(p-p_0)^2 \nonumber\\
            &=&+e(p-p_0)(q-q_0) + f(q-q_0)^2,
\end{eqnarray}
where $p=m_b/m_{\tau},p_0=4.75$~GeV$/1.777$~GeV$=2.6730,q=m_c/m_b,q_0=0.28$
and the polynomial coefficients can be fitted for each of the $\delta 
^{(i)}_n$ separately with a relative error of less than 2\%. 
Our choice of the central values reflects the realistic masses
of quarks: $m_b=4.75$~GeV and $m_c=1.35$~GeV,for which $\delta ^{(i)}_n=a_n^{(i)}$.
\par
To bring out the difference in the extent to which 
the corrections affect the two different
polarization states it is useful to compare these coefficients with
the ones obtained with the polarization summed over. This can be done
along the lines suggested by the treatment of the polarization itself,
see Eq.(\ref{PviaR}). The corresponding expansion takes the form
\begin{equation}
{{r_n^-}\over{r_n}}={{r^{-(0)}_n}\over{r_n^{(0)}}}\left[1-{{2\alpha_s}\over{3\pi}}(\delta _n^{(p)-}-\delta _n^{(p)})
+{{\lambda _1}\over{m_b^2}}(\delta _n^{(1)-}-\delta _n^{(1)})
+{{\lambda _2}\over{m_b^2}}(\delta _n^{(2)-}-\delta _n^{(2)})
\right].
\end{equation}
With these, one can readily find the actual relative correction of each kind, assuming
reasonable values of $\alpha _s,\lambda _1$~and$\lambda _2$. Here we take
$\alpha _s=0.2$, $0.15$~GeV$^2 \leq -\lambda_1 \leq 0.60$~GeV$^2$,
$\lambda_2=0.12$~GeV$^2$,
keep the $b$~quark mass fixed at $4.75$~GeV and the mass of the $c$ quark
equal to $1.35$~GeV. The corrections then read for $n=1(n=5)$
\begin{center}
\begin{tabular}{|c|c|c|}                       \hline
        & \multicolumn{2}{c|} {Correction to} \\ \hline
Type           & $r_n^-/r_n^{(0)-} $                & $r_n/r_n^{(0)}$
\\ \hline
perturbative   & $-0.00084(-0.0048)$                & $-0.0009(-0.0052)$
\\ \hline
kinetic energy & $0.008\pm 0.005(0.06\pm 0.04)  $   & $0.008\pm
0.005(0.06\pm 0.04)$ \\ \hline
chromomagnetic & $-0.0097(-0.053) $                 & $-0.0092(-0.0511)$
\\ \hline
\end{tabular}
\end{center}
\section{Summary}
The first order perturbative QCD correction to the polarization of
the charged lepton in semileptonic $B$ decays has been found
analytically. It is expressed by the charged lepton energy and
the invariant mass of the lepton system. The polarization correction has
turned out to be very small as it does not exceed $1\%$ taking
common values for the parameters occuring in the formulae. This makes
the polarization a very useful quantity in determining the quark masses.
The moments of $\tau$ lepton energy distribution have been evaluated
for the case of a polarized lepton and the correction has again
been found to be little different from the one in the case of an
unpolarized lepton. The nonperturbative HQET corrections to the
moments have also been calculated using the formulae from \cite{FLNN22}.

\end{document}